# Anisotropic magnon-magnon coupling in synthetic antiferromagnets

Wei He(何为)[1,*], Z. K. Xie(谢宗凯)[1], Rui Sun(孙瑞)[1], Meng Yang(杨萌)[1], Yang Li(李阳)[1], Xiao-Tian Zhao(赵晓天)[2,*], Wei Liu(刘伟)[2], Z. D. Zhang(张志东)[2], Jian-Wang Cai(蔡建旺)[1], Zhao-Hua Cheng(成昭华)[1], and Jie Lu(芦杰)[3,*]

[1]State Key Laboratory of Magnetism and Beijing National Laboratory for Condensed Matter Physics, Institute of Physics, Chinese Academy of Sciences, Beijing 100190, PR China

[2]Shenyang National Laboratory for Materials Science, Institute of Metal Research, Chinese Academy of Sciences, Shenyang, PR China

[3]College of Physics and Hebei Advanced Thin Films Laboratory, Hebei Normal University, Shijiazhuang 050024, PR China

**ABSTRACT**

  The magnon-magnon coupling in synthetic antiferromagnets advances it as hybrid magnonic systems to explore the quantum information technologies. To induce the magnon-magnon coupling, the parity symmetry between two magnetization needs to be broken. Here we experimentally demonstrate a convenient method to break the parity symmetry by the asymmetric structure. We successfully introduce a magnon-magnon coupling in Ir-based synthetic antiferromagnets CoFeB(10 nm)/Ir($t_{Ir}$=0.6 nm, 1.2 nm)/CoFeB(13 nm). Remarkably, we find that the weakly uniaxial anisotropy field (~ 20 Oe) makes the magnon-magnon coupling anisotropic. The coupling strength presented by a characteristic anticrossing gap varies in the range between 0.54 GHz and 0.90 GHz for $t_{Ir}$ =0.6 nm, and between 0.09 GHz and 1.4 GHz for $t_{Ir}$ = 1.2 nm, respectively. Our results demonstrate a feasible way to induce the magnon-magnon coupling by an asymmetric structure and tune the coupling strength by varying the direction of in-plane magnetic field. The magnon-magnon coupling in this highly tunable material system could open exciting perspectives for exploring quantum-mechanical coupling phenomena.

**Corresponding authors**:
E-mail: hewei@iphy.ac.cn
E-mail: xtzhao@imr.ac.cn
E-mail: jlu@yzu.edu.cn

Ferromagnetic layers coupled by interlayer exchange display a remarkable variety of modern magnetism. Especially, the synthetic antiferromagnet (SAF), which is an antiferromagnetic-coupled ferromagnet/nonmagnet/ferromagnet trilayer, has become promising nanotechnology in spintronics.[1–3] Usually, the interlayer exchange coupling (IEC) between the two ferromagnets is mediated by the nonmagnet through the Ruderman-Kittel-Kasuya-Yosida (RKKY) interaction.[4] Similar to the antiferromagnet, the advances in high frequency resonance and a small stray field make SAF a capable candidate for realizing high-performance magnetic memory devices.[5–7] Duo to the antiferromagnetic IEC, two magnetization precession modes, acoustic mode (AM, in-phase) and optical mode (OM, out-of-phase), are formed and have been explored in lots of SAFs.[8–13] Typically in the practical case of zero field, the resonance frequency of OM in SAF is higher than the acoustic mode[2] and can be tuned up to 20 GHz.[14] When applying a magnetic field in a certain range, the OM frequency reduces and AM frequency increases.[12] Once two modes approaching each other, they can be hybridized by the arising of the magnon-magnon coupling in SAF, which is analogous to the hybrid quantum system.[15,16] Since with the advantage of easy-fabrication and high-tunability, SAF as a host of magnon-magnon coupling is a good platform for hybrid magnonic systems to explore quantum information technologies.[17]

Recently, some progresses have been achieved at the hybridization of these two distinct modes in SAF.[15,16,18] Generally, the presence of parity or exchange symmetry protects mode crossings between AM and OM branches of FMR spectra,[19,20] which is very common in symmetrical SAF.[15,16] In fact, this mode crossing can be eliminated by inducing the symmetry breaking in SAF. One of the effective ways is to tilt the field toward out-of-plane direction, which will lift the system's rotation symmetry axis away from in-plane, thus breaks the rotation symmetry of the hard axis (out-of-plane) from the magnetostatic interaction.[15,20] In this case, a gap of 1.0 GHz was achieved in CoFeB(3 nm)/Ru (0.5 nm)/CoFeB(3 nm).[15] Another way recently reported is through the dynamic dipolar interaction (nonuniform excitation) to induce a gap of 0.67 GHz in CoFeB(15 nm)/Ru (0.6 nm)/CoFeB(15 nm).[16] In this case, when the field is tilted from the spin-wave propagation, the parity symmetry is broken. Therefore, the introduction of symmetry breaking is necessary for searching new approaches to induce magnon-magnon coupling in SAF.

On the other hand, an alternative approach has been discussed to induce the magnon-magnon coupling in the term of modifying the damping of a selective layer through the spin Hall effect.[18] From a device perspective, the inducing magnon-magnon interaction in a SAF trilayer is more attractive when it is compatible with the spin-orbitronic devices. Ru is the most used material as the interlayer of SAF since it can provide a large coupling strength and be easy to tune the RKKY coupling by the thickness.[10] However, it lacks good spin-orbit properties. On the other hand, Ir also can induce large IEC in SAF.[6,21,22] More, it has a large spin-orbit coupling. It has the advantages to modify the damping[23], induce PMA[7] and Dzyaloshinskii-Moriya interaction[24,25], generate spin-orbit torque[7], and so on. Ir-based SAF enables us to build a hybrid magnonic platform integrating the advantages of spin-orbitronic devices. It could offer more opportunities to tailor the coupling phenomena. So far, to realize the magnon-magnon coupling in Ir-based SAF is feasible and highly desirable.

In this letter, we report the observation of strong and anisotropic magnon-magnon coupling in Ir-based SAF CoFeB(10 nm)/Ir($t_{Ir}$)/CoFeB(13 nm), through the broadband ferromagnetic resonance (FMR), where $t_{Ir}$ presents the thickness of Ir layer. Due to the asymmetric thickness of two magnetic layers, the IEC fields are different for two layers and then break the parity symmetry of two magnetizations as well as introduce a magnon-magnon coupling. In two selected samples with $t_{Ir}$ =0.6 nm (the strongest IEC) and 1.2 nm (the second strongest IEC), the magnon-magnon coupling was explored and determined. More, when rotating the sample in in-plane, we found that the magnon-magnon coupling is anisotropic. The coupling strength presented by a characteristic anticrossing gap varies in the range between 0.54 GHz and 0.90 GHz for $t_{Ir}$ =0.6 nm, and between 0.09 GHz and 1.40 GHz, for $t_{Ir}$ =1.2 nm, respectively. The magnon-magnon coupling in this highly tunable material system could offer more opportunities to tailor the coupling phenomena.

The schematic diagram of the multilayer stack is shown in Fig.1(a). A uniaxial anisotropy is always induced by some unavoidable issues during the film deposition by sputtering, i.e., the strain, the roughness of substrate, the oblique deposition, and so on. In our samples, we attribute it mainly to the oblique deposition. The static hysteresis loops of CoFeB were measured by using a vibrating sample magnetometer (VSM). Figs. 1(a)-(f) show the normalized hysteresis loops of CoFeB/Ir/CoFeB for the easy and hard axis at varied thickness $t_{Ir}$ =0.5 nm, 0.6 nm, and 1.2 nm, respectively. The loops demonstrate the varied interlayer coupling with increasing Ir thickness. The curve in Fig. 1(a) is a squared loop and in Fig. 1(b) is a loop with little hysteresis. We consider the sample with $t_{Ir}$ =0.5 nm as the uncoupled or weak ferromagnetic coupled trilayer with a uniaxial magnetic anisotropy. However, the loops have big changes when $t_{Ir}$ is 0.6 nm (as shown in Fig. 1(d) and Fig. 1(d)). In this sample, when the magnetic field decreases from the saturated state, the magnetization decreases until a plateau is reached. It reveals the two CoFeB layers antiparallel caused by the antiferromagnetic IEC. The finite remnant for this plateau is due to the difference of thicknesses for two CoFeB layers. The saturated fields for the easy axis (Fig. 1(c)) and hard axis (Fig. (d)) are around 1300 Oe. With increasing the thickness $t_{Ir}$ to 0.7 nm, the loops (not shown here) change back to an uncoupled or weak ferromagnetic-coupled state, like $t_{Ir}$ =0.5 nm. The state is kept until $t_{Ir}$ =1.1 nm, the antiferromagnetic coupling rises again and keeps in $t_{Ir}$ =1.1~1.4 nm. As a typical example for this period, the loops for $t_{Ir}$ =1.2 nm are plotted in Fig. 1(e) and Fig. 1(f). The saturated fields for the easy axis (Fig. 1(e)) and hard axis (Fig. (f)) are 90 Oe and 123 Oe, respectively. It indicates that this interlayer coupling strength is less than one of $t_{Ir}$ =0.6 nm.

In order to quantitatively determine the strength of IEC, the ferromagnetic resonance frequency was used rather than the saturated field which was used in the previous reports.[5,6,26] Fig. 1(g) shows the frequency power spectra of zero field as a function of Ir thickness $t_{Ir}$. The resonance peak appears in the spectra and its location in the frequency axis is varied in an oscillating behavior, which is related to the characteristic RKKY interaction. Furthermore, the ferromagnetic resonance frequency $f_{res}$ was counted by the peaks of spectra and were plotted in Fig. 1(h) also as the function of Ir thickness. The zero field resonance frequency $f_{res}$ of OM is described for the antiferromagnetic IEC by the following equation as[27,28]

$$f_{res}^o = \frac{\gamma}{2\pi}\sqrt{(H_u + 2H_{ex1} - 4H_{ex2})(H_u + 4\pi M_s)} \ . \tag{1}$$

Here, $\gamma/2\pi$ is 28 GHz/T, and $4\pi M_s$ is 1.4 T from the VSM measurement, $H_u$ is the in-plane uniaxial anisotropy field, and $H_{ex1}$ and $H_{ex2}$ are the effective bilinear and biquadratic exchange coupling fields, respectively (see more description of the effective fields in the Supporting Information). Once IEC vanishes, the resonance frequency at zero field becomes $\frac{\gamma}{2\pi}\sqrt{H_u(H_u + 4\pi M_s)}$ as the well-known uniform mode. To simply clarify the variation of IEC, the $H_u$ is selected as 24 Oe obtaining from $t_{Ir}$ =0.7 nm and then used for all samples. The whole of $2H_{ex1} - 4H_{ex2}$ treated as the fitting parameter was obtained for varied Ir thickness. The interlayer exchange coupling field $2H_{ex1} - 4H_{ex2}$ is obtained and plotted in Fig. 1(h). It is clear that the strength of IEC shows an oscillation period of 0.6 nm. It is a typical character of RKKY. The first RKKY peak appears at $t_{Ir}$ =0.6 nm and the second appears at $t_{Ir}$ =1.2 nm. Usually, $H_{ex2}$ is neglectable in the discussion. It is true for our samples except for $t_{Ir}$ =0.6 nm. The IEC field $2H_{ex1} - 4H_{ex2}$ is 550 Oe for $t_{Ir}$ =0.6 nm, which is smaller than the saturated field $H_s \approx$ 1300 Oe. As known, $H_s$ is $2H_{ex1} + 4H_{ex2} - H_u$. In such a case, we have to consider the contribution of the biquadratic exchange coupling in the IEC.

The ferromagnetic resonance spectra are mapped as the function of the external field along the easy axis and hard axis in Fig. 2(a) and Fig. 2(b) for the sample with $t_{Ir}$ =0.6 nm, in Fig. 2(c) and Fig. 2(d) for the sample with $t_{Ir}$ =1.2 nm, respectively. To better illustrate the peaks of the two branches, the FMR signal is presented by real part of $S_{21}$. All of them include two branches: OM and AM. When the field approaching zero field, the two layers are antiferromagnetic coupling and the net magnetization is too small to observe AM signal. This is also the reason why only OM displays in Fig. 1(g) for samples with the presence of IEC. With increasing field, OM in easy axis is nearly constant with little increase until the field larger than 140 Oe, the critical field for spin-flop. After spin-flop, the magnetizations of two layers change continuously from noncollinear to parallel. For the hard axis, it is similar to the case of the easy axis after spin-flop. Interestingly, an anticrossing gap appears between AM and OM. The gap is counted as the minimum difference between high and low frequencies branches. The FMR curves at selected fields (the data from Fig.2(a)) are plotted in Fig. 2(e). The minimum gap is located in the FMR curver of 747 Oe, which is marked as a strengthened red line. The gap reflects the strength of magnon-magnon coupling. The counted gap for $t_{Ir}$ =0.6 nm is 0.64 GHz and 0.54 GHz for the easy and hard axis, respectively. For $t_{Ir}$ =1.2 nm, the gap is 0.62 GHz for the easy axis and disappears at the hard axis. The anisotropic gap demonstrates the possibility of tunning the magnon-magnon interaction in SAF by varying the direction of the external magnetic field.

To understand magnon-magnon coupling quantitively, a more complete model is employed to numerically calculate the FMR spectra as a function of the external field (See the Supporting Information). The parameters in the model include the in-plane uniaxial anisotropy field $H_u$, the demagnetization field $4\pi M_s$, and the effective bilinear exchange coupling field $H_{ex1}$ as well as the biquadratic exchange coupling field $H_{ex2}$. The first two parameters are same for two-layers and the last two are thickness-dependent for two layers (See the Supporting Information).[28] The optimal calculated dispersion curves for both AM and OM are also plotted in Fig. 2(a)-(d). The used and calculated parameters are listed in Table I. As seen, the calculated curves replicate the dispersion of OM and AM quite well for both two samples as well as the field along the easy or hard axis. For $t_{Ir}$ =0.6 nm, $H_{ex2}:H_{ex1} \approx 0.2$. The biquadratic exchange coupling is seldom reported and determined in Ir-based SAF. The emergence of larger biquadratic exchange coupling indicates the complexity of RKKY interaction in Ir. Further work is needed to treat it in the view of electronic band structure or roughness.[29] Remarkably, the anticrossing gaps are also replicated as well as their anisotropy in two directions (see Table I). The appearance of an anticrossing gap is the result of mutual magnon-magnon interaction between AM and OM, which indicates the presence of symmetry breaking in our samples. As discussed above, the obliquely-oriented external field[15] or the dynamical dipolar interaction[16] has been employed for the symmetry breaking. In our experiment, the samples are asymmetrical SAFs with different thickness of magnetic layers. Fig.2(f) briefly illustrates this mechanism. The IEC field is proportional to the inverse of thickness. For symmetric structure, the IEC field is same for two layers. The magnetization of two layers satisfies the two-fold rotational symmetry $C_2$ along the axis of external field. For asymmetric structure, the effective IEC fields are not equal. Then $C_2$ symmetry is broken when two magnetizations are neither parallel nor antiparallel. This approach of symmetry breaking has been discussed in detail on how to induce magnon-magnon coupling, which was summarized in ref.[20]

To explore the mechanism of the anisotropic magnon-magnon coupling in our experiment, the ferromagnetic resonance spectra were acquired at varying the magnetic field direction in-plane. The FMR spectra at varying field direction for $t_{Ir}$ =1.2 nm are presented as the example in Fig. S1 (see Supporting Information). The values of the counted gap as the function of $\theta_H$ are plotted in Fig. 3(a) and 3(b) for $t_{Ir}$ =0.6 nm and $t_{Ir}$ =1.2 nm, respectively. Here, $\theta_H$=0º and 90º present the easy and hard axis, respectively. The maximum is 0.9 GHz at $\theta_H$=36º for $t_{Ir}$ =0.6 nm and 1.4 GHz at $\theta_H$=36º for $t_{Ir}$ =1.2 nm. Based on the method of resolving FMR spectra (see Supporting Information) and the parameters in Table I, the calculated values are also plotted with and without $K_u$. There is a quantitative agreement between experimental data and the calculated curves in Fig. 3(a) and 3(b). The marginal deviation at a low angle for $t_{Ir}$ =1.2 nm is the result of the disturbance (see description in Supporting Information). We note that the predicated gap for $t_{Ir}$ =1.2 nm at the hard axis is 0.09 GHz, which is too small to be distinguished since the linewidth makes two branches overlap. In comparison with the calculated curves with and without $K_u$, we can clarify the anisotropy of magnon-magnon coupling resulted from the uniaxial anisotropy. It was reported in a compensated ferrimagnet GdIG that a weakly magnetocrystalline anisotropy can lead an anisotropic magnon-magnon coupling, since it can break rotational symmetry.[30] In our sample, the parity-symmetry is broken firstly by asymmetric IEC fields, then the symmetry-breaking can further enhance or suppress by the uniaxial anisotropy. The anisotropy of the gap is larger for $t_{Ir}$ =1.2 nm when counting the ratio of maximum vs minimum of gap. More, for $t_{Ir}$ =1.2 nm, the gap is enhanced more largely up to 1.4 GHz. Even we can calculate it numerically, the work of theoretical analysis will be done in the future to discuss the relationship among the strength of IEC, $K_u$, and the anisotropic magnon-magnon coupling. Whatever, the uniaxial field is a feasible parameter to modulate the magnon-magnon coupling for other SAF and to obtain strong magnon-magnon coupling even with a weak IEC.

Summary, we used FMR to determine the strength of IEC as well as explore the magnon-magnon interaction in Ir-based SAF: CoFeB(10 nm)/Ir($t_{Ir}$)/CoFeB(13 nm). The IEC has an oscillation period of 0.6 nm, while the first peak appears at $t_{Ir}$ =0.6 nm and the second peak appears at $t_{Ir}$ =1.2 nm. The biquadratic interlayer coupling was determined for $t_{Ir}$ =0.6 nm as larger as one-fifth of bilinear interlayer coupling. Remarkably, we observed the magnon-magnon coupling in these SAF trilayers. The asymmetry IEC fields benefited from asymmetry thickness is the key issue to induce symmetry breaking and then magnon-magnon coupling. More, the weakly uniaxial anisotropy field (~ 20 Oe) makes the magnon-magnon coupling anisotropic. The coupling strength varies in the range between 0.54 GHz and 0.90 GHz for $t_{Ir}$ =0.6 nm, and between nearly zero to 1.4 GHz for $t_{Ir}$ = 1.2 nm, respectively. This offers the tunability of magnon-magnon interaction by varying the direction of the external magnetic field in in-plane. Our results demonstrate a feasible way to induce the magnon-magnon coupling by an asymmetric structure in SAF and enable the tuning it by varying the direction of the magnetic field in in-plane. So far, these strategies can be employed for other SAF. We emphasize that Ir-based SAF enables us to build a hybrid magnonic platform integrating the advantages of spin-orbitronic devices. It could offer more opportunities to tailor the coupling phenomena.

**Experimental details**

**Sample fabrication**: Ta/CoFeB(10 nm)/Ir($t_{Ir}$)/CoFeB(13 nm)/Ta films were deposited on thermally oxidized silicon wafer by dc magnetron sputtering at an Ar pressure of 0.5 Pa. The thicknesses for the buffer and capping Ta layers are both 3 nm. The two magnetic layers with thicknesses of 10 nm and 13 nm form an asymmetry spin-valve structure are beneficial to induce the magnon-magnon interaction. For the Ir layer, thickness $t_{Ir}$ is varied from 0.5 nm to 1.4 nm. The base pressure of the sputtering system was less than $4\times10^{-5}$ Pa. Sputtering rates for CoFeB, Ir, and Ta were about 0.37, 0.18, and 0.27 Å/s, respectively.

**FMR Measurements**: The dynamical properties were explored by the broadband ferromagnetic resonance (FMR). The FMR measurements were performed in a vector network analyzer (VNA). The FMR setup is illustrated in Fig. S2. The sample on coplanar waveguide(CPW) is 5mm×5mm. The *S*-line of CPW is 0.2 mm. The film was positioned on the waveguide by flipping the sample. The VNA-FMR was operated in the frequency-swept mode (1-17 GHz) and the static magnetic field was applied in the film plane. The microwave power is set at 1 mW (0 dBm). The FMR was carried out by sweeping the frequency and the transmission parameters $S_{21}$ was acquired as a function of microwave frequency for a series of fixed magnetic fields. To remove the nonmagnetic response contribution, a suitable microwave calibration was carried out.[31] Then $S_{21}$ only derived from the magnetic response was obtained. It can be expressed as $S_{21} = -iAf\chi(f)e^{i\phi}$, A is a scaling parameter that is proportional to the coupling between sample and waveguide, $f$ is the microwave frequency, $\phi$ is the phase delay, and $\chi$ is the magnetic susceptibility, respectively.[31] All results presented in this paper were obtained at room temperature.

**Supporting information**

Supporting Information is available.

> Additional information about the decription of the energy model; detailed information about the solution of ferromagnetic resonance in SAF; the FMR spectra for $t_{Ir}$ =1.2 nm at varying $\theta_H$; the setup of VNA-FMR.

**ACKNOWLEDGEMENTS**

This project was supported by the National Natural Sciences Foundation of China (Grant Nos. 51871235, 51671212, 52031014, 51771198, and 51801212), the National Key Research Program of China (Grant Nos. 2016YFA0300701, 2017YFB0702702, and 2017YA0206302), and the Key Research Program of Frontier Sciences, CAS (Grant Nos. QYZDJ-SSW-JSC023, KJZD-SW-M01, and ZDYZ2012-2). J.L. acknowledges support from the Natural Science Foundation for Distinguished Young Scholars of Hebei Province of China (A2019205310).

Table I. The parameters for CoFeB(10 nm)/Ir($t_{Ir}$)/CoFeB(13 nm) for $t_{Ir}$=0.6 nm and 1.2 nm, respectively. The magnetization was measured by a vibration sample magnetometer. The in-plane uniaxial anisotropy field $H_u$ is the same for two layers. The effective bilinear exchange coupling field $H_{ex1}$ and biquadratic exchange coupling field $H_{ex2}$ are thickness-dependent, here we denote them for 10 nm CoFeB. The gaps are counted in both of easy and hard axis. The experimental (exp) and calculated (cal) values are both presented.

| $t_{Ir}$ (nm) | $4\pi M_s$ (T) | $H_u$ (Oe) | $H_{ex1}$ (Oe) | $H_{ex2}$ (Oe) | Gap (GHz) Easy | Hard |
|---|---|---|---|---|---|---|
| 0.6 | 1.4 | 20 | 480 | 95 | exp: 0.64 | 0.54 |
| | | | | | cal: 0.67 | 0.58 |
| 1.2 | 1.4 | 20 | 54 | 0 | exp: 0.62 | 0 |
| | | | | | cal: 0.71 | 0.09 |

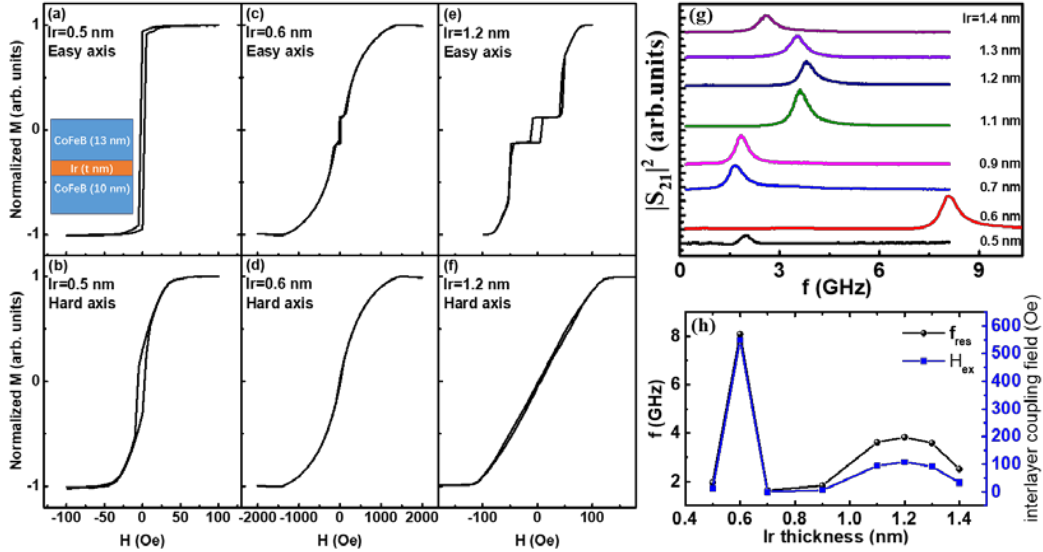

**Fig 1.** (color online) Normalized hysteresis loop with external field along in-plane easy axis and hard axis for Ir thickness t=0.5 nm (a) and (b), t=0.6 nm (c) and (d), and t=1.2 nm (e) and (f), respectively. (g) The broadband frequency power spectra $|S_{21}|^2$ at zero field as the varied thickness of Ir layer. (h) The counted resonance frequency from the spectra in (g) and the estimated interlayer exchange coupling field as a function of Ir thickness. The insert in (a) is the schematic diagram of the multilayer stack CoFeB(10 nm)/Ir/CoFeB(13 nm).

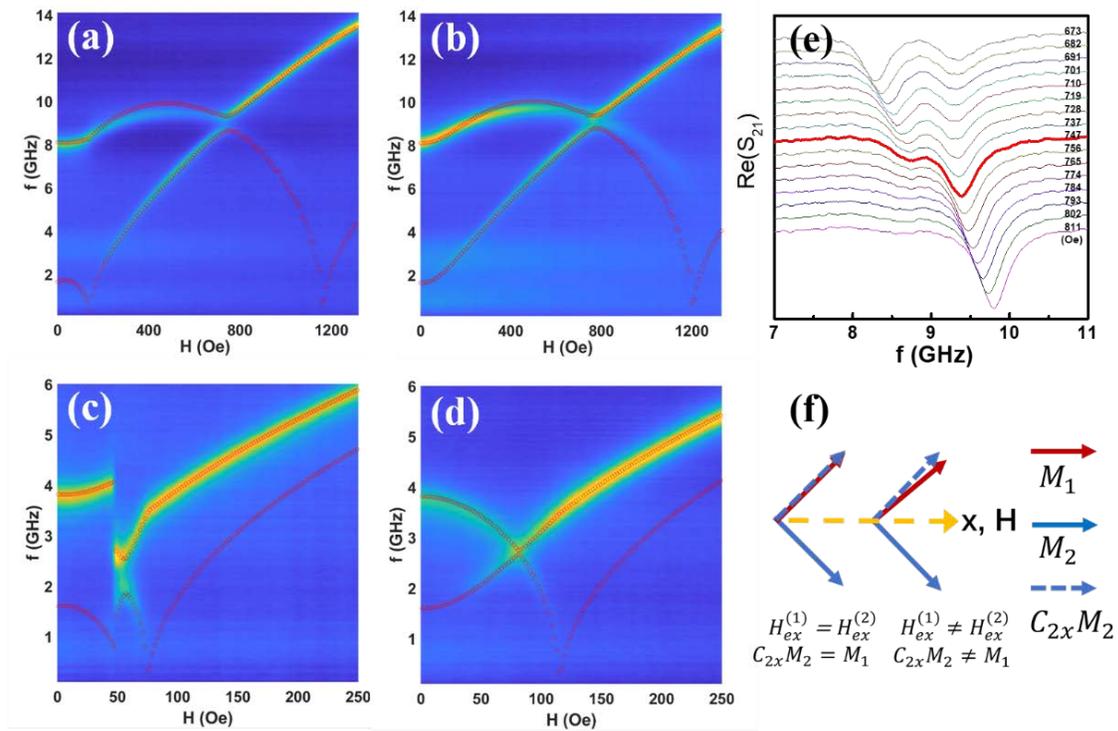

**Figure 2** (color online) Ferromagnetic resonance spectra presented by the real part of $S_{21}$ mapping as the function of the external field along and in (a) easy axis and (b) hard axis for the sample with $t_{Ir}$ =0.6 nm, in (c) easy axis and (d) hard axis for the sample with $t_{Ir}$ =1.2 nm, respectively. The red circular lines in Fig. 3(a)-(d) are the optimal calculated dispersion curves for both AM and OM. (e) The FMR curves at selected fields extracted from Fig. 2(a). The FMR curve for 747 Oe is marked as a strengthened red line to illustrate the anticrossing gap. (f) The diagram for demonstrating rotational symmetry in symmetric structure and symmetry breaking in asymmetric structure. $C_{2x}$ is the 2-fold rotational operator around the x-axis or external field. For further information see the content.

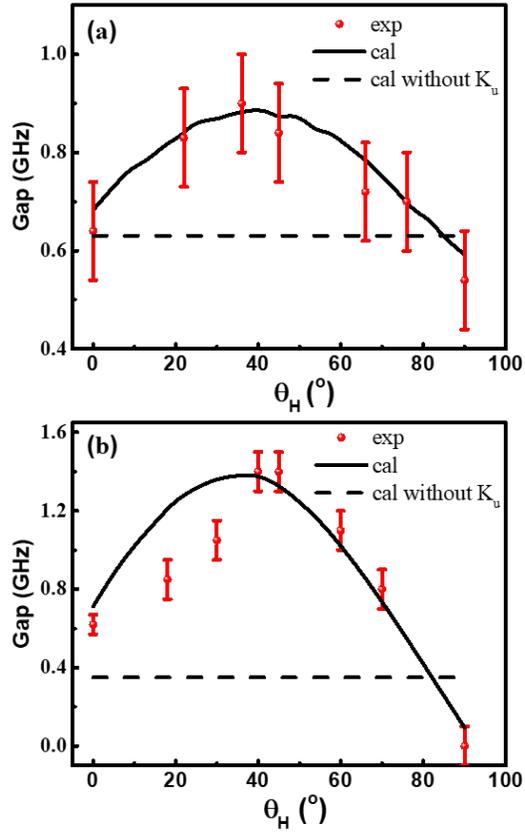

**Figure 3** (color online) The values of counted gap as the function of θ$_H$ for (a) $t_{Ir}$ =0.6 nm and (b) $t_{Ir}$ =1.2 nm, respectively. Here, θ$_H$=0° and 90° present the easy and hard axis, respectively. The calculated values are based on the method (see Supporting Information) and the parameters in Table I with (strength lines) and without $K_u$ (dash lines).

# Supporting Information
# Anisotropic magnon-magnon coupling in synthetic antiferromagnets

Wei He(何为)[1,*], Z. K. Xie(谢宗凯)[1], Rui Sun(孙瑞)[1], Meng Yang(杨萌)[1], Yang Li(李阳)[1], Xiao-Tian Zhao(赵晓天)[2,*], Wei Liu(刘伟)[2], Z. D. Zhang(张志东)[2], Jian-Wang Cai(蔡建旺)[1], Zhao-Hua Cheng(成昭华)[1], and Jie Lu(芦杰)[3,*]

[1]State Key Laboratory of Magnetism and Beijing National Laboratory for Condensed Matter Physics, Institute of Physics, Chinese Academy of Sciences, Beijing 100190, PR China

[2]Shenyang National Laboratory for Materials Science, Institute of Metal Research, Chinese Academy of Sciences, Shenyang, PR China

[3]College of Physics and Hebei Advanced Thin Films Laboratory, Hebei Normal University, Shijiazhuang 050024, PR China

**Corresponding authors**:

E-mail: hewei@iphy.ac.cn

E-mail: xtzhao@imr.ac.cn

E-mail: jlu@yzu.edu.cn

## 1. The description of the energy model

For the case when the external magnetic field $H$ is applied in the plane of films, the energy for the system including two identical ferromagnetic layers is[1]

$$E = \sum_{i=1}^{2} d_i[-M_i H\cos(\theta_i - \theta_H) - K_u \cos^2 \theta_i] + J_1 \frac{\mathbf{M_1} \cdot \mathbf{M_2}}{M_1 M_2} + J_2 \left(\frac{\mathbf{M_1} \cdot \mathbf{M_2}}{M_1 M_2}\right)^2.$$

where $M_1 = M_2 = M_s$ is the magnetization for each CoFeB layer, and $d_i$ ($i=1,2$) is the thickness for the top and bottom layer, i.e., $d_1$ is 12 nm and $d_2$ is 10 nm. $K_u$ is the uniaxial in-plane anisotropy. Its easy axis is defined as the x-axis. $\theta_i$ and $\theta_H$ are defined as the angle of equilibrium directions for two magnetizations and the external field with respect to the x-axis, respectively. Here, both the so-called bilinear coupling constant $J_1$ and biquadratic coupling constant $J_2$ are included. The equilibrium configuration ($\theta_1$ and $\theta_2$) can be obtained through find the minimum of energy. At zero field, if $J_1$ is positive, the bilinear coupling favors two magnetic layers antiparallel. It is the so-called antiferromagnetic coupling. On other hand, $J_2$ is positive, the biquadratic coupling favors the two magnetizations 90°-type coupling. More cases have been discussed in ref.[2].

## 2. The solution of ferromagnetic resonance in SAF

For convenience, we define an effective anisotropy field and effective exchange fields as $H_u = 2K_u/M$, $H_{ex1}^{(i)} = \frac{J_{1,i}}{Md_i}$, and $H_{ex2}^{(i)} = J_{2,i}/Md_i$, $i=1,2$ for the top and bottom magnetic layer, respectively. Based on the coupled Landau-Lifshitz equation (LL), the resonance frequency of magnetization precession can be treated as the solution of the linearized LL equation in the case of a small amplitude of magnetization precession. The dispersion is numerical determined as the eigenvalue of the following matrix:[1]

$$\begin{bmatrix} 0 & H_1 & 0 & H_2 \\ -H_3 & 0 & H_4 & 0 \\ 0 & H_6 & 0 & H_5 \\ H_8 & 0 & -H_7 & 0 \end{bmatrix}.$$

Where

$$H_1 = H\cos(\theta_1 - \theta_H) + H_u \cos 2(\theta_1 - \theta_u) + 4\pi M_s + H_{ex1}^{(1)} \cos(\theta_1 - \theta_2)$$
$$- 2H_{ex2}^{(1)} \cos 2(\theta_1 - \theta_2)$$

$$H_2 = -H_{ex1}^{(2)} + 2H_{ex2}^{(2)} \cos(\theta_1 - \theta_2)$$

$$H_3 = H\cos(\theta_1 - \theta_H) + H_u \cos[2(\theta_1 - \theta_u)] + H_{ex1}^{(1)} \cos(\theta_1 - \theta_2)$$
$$- 2H_{ex2}^{(1)} \cos[2(\theta_1 - \theta_2)]$$

$$H_4 = H_{ex1}^{(2)} \cos(\theta_1 - \theta_2) - 2H_{ex2}^{(2)} \cos[2(\theta_1 - \theta_2)]$$

$$H_5 = H\cos(\theta_2 - \theta_H) + H_u \cos 2(\theta_2 - \theta_u) + 4\pi M_s + H_{ex1}^{(2)} \cos(\theta_1 - \theta_2)$$
$$- 2H_{ex2}^{(2)} \cos 2(\theta_1 - \theta_2)$$

$$H_6 = -H_{ex1}^{(1)} + 2H_{ex2}^{(1)} \cos(\theta_1 - \theta_2)$$

$$H_7 = H\cos(\theta_2 - \theta_H) + H_u \cos[2(\theta_2 - \theta_u)] + H_{ex1}^{(2)} \cos(\theta_1 - \theta_2)$$
$$- 2H_{ex1}^{(2)} \cos[2(\theta_1 - \theta_2)]$$

$$H_8 = H_{ex1}^{(1)} \cos(\theta_1 - \theta_2) - 2H_{ex2}^{(1)} \cos[2(\theta_1 - \theta_2)]$$

The eigenvalue is $-i\omega/\gamma = -i2\pi f_r/\gamma$.

### 3. The FMR spectra for $t_{Ir}$ =1.2 nm at varying $\theta_H$

The ferromagnetic resonance spectra were acquired at varying the magnetic field direction $\theta_H$. Since the anisotropy of the gap is larger in the sample with $t_{Ir}$ =1.2 nm, the spectra for $t_{Ir}$ =1.2 nm is presented in Fig. S1. The red circle curves are calculated based on the parameters from Table I and are also plotted in Fig. S1. At low angles like $\theta_H$=18° and 30º (see Fig. S1(a) and (b)), the location of the anticrossing gap closes to the spin-flop field. After spin-flop, the magnetization is not aligned uniformly immediately but after a field window since the broaden distribution of IEC and $H_u$. It brings the disturbance for counted the gap. Therefore, a significant deviation happens due to this non-uniform.

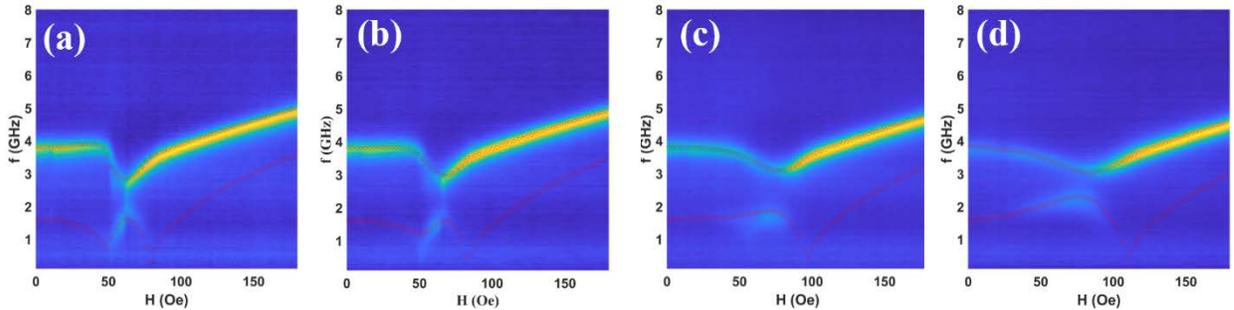

**Fig. S1**. FMR spectra for CoFeB(10 nm)/Ir(1.2 nm)/CoFeB(13 nm) at varying $\theta_H$, (a) $\theta_H=18º$, (b), $\theta_H=30º$, (c) $\theta_H=45º$, (b), $\theta_H=70º$, respectively. The red circle curves are calculated based on the parameters from Table I.

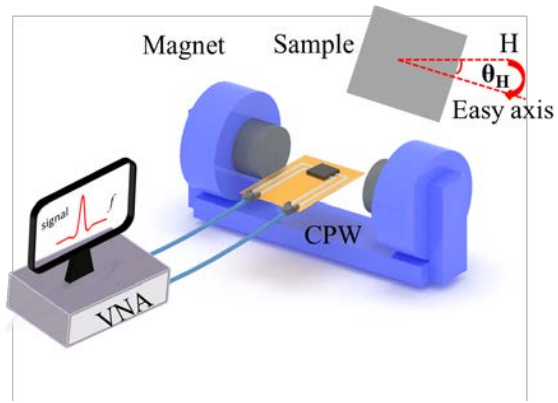

**Fig. S2**. The illustration of VNA-FMR setup. The sample is rotated in in-plane. $\theta_H$ is the angle between its easy axis and the external field.